\documentclass[pdftext,a4wide,12pt]{article}
\usepackage[dvips]{graphicx}
\usepackage{epsfig}
\usepackage{amsfonts}
\usepackage{graphics,color}
\begin{document}
\def\ti{\tilde}
\def\da{\dagger}
\def\a{\alpha}
\def\b{\beta}
\def\g{\gamma}
\def\l{\lambda}
\def\d{\delta}
\def\k{\kappa}
\def\p{\partial}
\def\t{\theta}
\def\s{\sigma}
\def\G{\Gamma}
\def\ap{\approx}
\def\v{\varepsilon}
\def\tr{\textrm}
\def\nn{\nonumber}
\def\w{\wedge}
\def\hw{\hat{\omega}}
\def\K{K\H{a}hler}
\def\o{\omega}
\def\na{\nabla}
\def\t{\theta}
\def\s{\sigma}
\def\G{\Gamma}
\def\D{\Delta}
\def\v{\varepsilon}
\def\tr{\textrm}
\def\nn{\nonumber}
\def\w{\wedge}
\def\hw{\hat{\omega}}
\def\K{K\H{a}hler}
\def\o{\omega}
\newcommand{\be}{\begin{eqnarray}}
\newcommand{\ee}{\end{eqnarray}}
\newcommand{\beq}{\begin{equation}}
\newcommand{\eeq}{\end{equation}}
\newcommand{\rc}{\nonumber\\}

\pagenumbering{arabic}
\renewcommand{\theequation}{\thesection.\arabic{equation}}

\definecolor{M_Beige}         {rgb}{0.96 , 0.96 , 0.86}
\newcommand{\CBei}[1]{\textcolor{M_Beige}{#1}}
\definecolor{M_Brown}         {rgb}{0.65 , 0.16 , 0.16}
\definecolor{M_Gold}          {rgb}{0.12 , 0.84 , 0.30}
\newcommand{\CGol}[1]{\textcolor{M_Gold}{#1}}
\definecolor{M_LemonChiffon}  {rgb}{1.00 , 0.98 , 0.80}
\newcommand{\CLem}[1]{\textcolor{M_LemonChiffon}{#1}}
\definecolor{M_Orange}        {rgb}{1.00 , 0.60 , 0.00}
\newcommand{\Ccya}[1]{\textcolor{M_Orange}{#1}}
\definecolor{M_Pink}          {rgb}{1.00 , 0.75 , 0.80}
\newcommand{\CPin}[1]{\textcolor{M_Pink}{#1}}
\definecolor{M_Gre}          {rgb}{0.05 ,0.46 , 0.00}
\newcommand{\Cred}[1]{\textcolor{M_Gre}{#1}}

\newcommand{\Cbla}[1]{\textcolor{black}{#1}}
\newcommand{\Cblu}[1]{\textcolor{blue}{#1}}
\newcommand{\Cgre}[1]{\textcolor{red}{#1}}
\newcommand{\Cyel}[1]{\textcolor{yellow}{#1}}
\newcommand{\Cwhi}[1]{\textcolor{white}{#1}}

\begin{titlepage}
\begin{center} \Large{ \bf String Dual of ${\cal N}=1$ Super Yang-Mills on the Cylinder}

\end{center}

\vskip 0.3truein
\begin{center}
Felipe Canoura${}$
\footnote{canoura@fpaxp1.usc.es}

\vspace{0.3in}

${}$Departamento de F\'\i sica de Part\'\i culas, Universidade de
Santiago de Compostela \\and\\
Instituto Galego de F\'\i sica de Altas Enerx\'\i as (IGFAE)\\
E-15782 Santiago de Compostela, Spain

\end{center}
\vskip 1
truein

\begin{center}
\bf ABSTRACT
\end{center}

\noindent We study the stringy description of ${\cal N}=1$ supersymmetric $SU(N)$ gauge theory on ${\mathbb {R}}^{1,2}\times S^1$. Our description is based on the known Maldacena-N\'u\~nez solution, properly modified to account for the compact dimension. The most interesting of its properties is that extra BPS M-branes are present, which generate a non-perturbative superpotential that we explicitly compute. 

\end{titlepage}
\setcounter{footnote}{0}

\section{Introduction}

The duality between string and field theory, in the framework originated by the AdS/CFT correspondence \cite{Maldacena:1997re,Gubser:1998bc,Witten:1998qj}, provides powerful tools to investigate the strong coupling dynamics of the latter. In view of their rich and quite well understood dynamics, ${\cal N}=1$ supersymmetric Yang-Mills (SYM) theories provide perhaps the best example to be studied in this perspective. One of their nice properties is that, sometimes, their infrared strong coupling properties can be encoded in a superpotential sourced by non-perturbative effects. The non-perturbative generation of a superpotential makes them appealing also for cosmological purposes and the related moduli stabilization problem \cite{Giddings:2001yu,Kachru:2003aw,Kachru:2003sx}. In general it is important to find examples where such superpotential can be directly calculated.

In this talk we find such a concrete example by studying the string dual of ${\cal N}=1$ SYM on the cylinder (by cylinder we mean the flat space ${\mathbb R}^{1,2}\times S^1$, {\it i.e.} four-dimensional Minkowski space with one spatial direction compactified). This cylindrical geometry improves the understanding of the string theoretic description of non-perturbative gauge phenomena. From the field theory point of view, the generation of a superpotential in this geometry is nicely described  in \cite{Davies:1999uw}. 
 
Our analysis starts from the Maldacena-N\'u\~nez (MN) \cite{Maldacena:2000yy} solution, which describes the infrared (IR) of ${\cal N}=1$ pure SYM theory. Its ultraviolet (UV) completion is instead related to little string theory and the two regimes of the theory are not smoothly connected in terms of a unique solution (they are S-dual to each other). The source of this problem is the bad asymptotic behavior of the dilaton. On the gauge theory side this reflects the difficulties of joining the weak coupling with the strong coupling regime of confining SYM theory in a unifying picture.

On the field theory side it is known that such an interpolating picture exists if we compactify one spatial dimension and consider SYM on ${\mathbb R}^{1,2}\times S^1$ \cite{Davies:1999uw}. In this case the non-perturbative physics is much better understood and typically infrared phenomena (such as gaugino condensation) have a semiclassical exhaustive description  \cite{Davies:1999uw}. It is indeed possible to explicitly write a non-perturbatively generated superpotential that leads to  a mass gap (providing a mass for the ``magnetic«" photons) and gaugino condensation.
  
To investigate SYM theory on the cylinder, we look for the proper modification of the MN background. When one deals with compact directions (as in the cylinder geometry we are considering here) the natural thing to do is T-duality. We then T-dualize the IIB MN solution along one of its flat spatial directions and consider the corresponding type IIA solution. This could be enough to study SYM on the cylinder, but the dilaton still diverges. As it is well known, this is a sign of the opening of the eleventh dimension. We then uplift the solution to eleven dimensions and find a globally well behaved solution. In this set-up SYM theory is the theory living on the worldvolume of $N$ M5-branes that wrap a three cycle with topology $S^2\times S^1$. Their backreaction generates the dual background. The eleven dimensional solution encodes in a non-trivial way the information that the dual SYM theory has one compact spatial direction. 

In principle the solution we build in the way sketched above is just valid to describe the infrared of  ${\cal N}=1$ SYM. We find that  also its UV description (related to NS5-branes in type IIB) corresponds to the same eleven dimensional solution. We have then a unique picture connecting the UV and the IR of the gauge theory in terms of the worldvolume theory of $N$ M5-branes in the background we find. 

On this solution we perform various gauge theory computations and we find perfect agreement with expectations. Both for the perturbative and non-perturbative calculations, it is crucial to use the M5 and M2 worldvolume actions. In particular, we find that in this set-up the theory is naturally formulated in terms of the scalar field dual to the three dimensional vector and such dualization has not to be imposed, as it is usually done, by hand. Such degree of freedom is related to the ``self-dual'' three form living on the M5 worldvolume (to which the boundary of the M2-branes couples to). 

The main novelty is that in this solution new kinds of instantons (the so-called ``fractional instantons") have a natural description. They are responsible  for the generation of the non-perturbative superpotential that we compute. In the M-theory description, they correspond to Euclidean M2 branes wrapping a three-cycle. Precisely two zero-modes are left by such configurations. This is the right number to generate a non-perturbative contribution to the superpotential \cite{Witten:1996bn}.

\section{Eleven dimensional solution dual to Maldacena-N\'u\~ nez}
\label{sec11}

The Maldacena-Nu\~nez background is a solution of the equations of motion of
type IIB supergravity \cite{Maldacena:2000yy}. If we perform a T-duality along one of the Minkowski coordinates and call $z$ to the periodic T-dual coordinate, the type IIA background we obtain is (in string frame):
\begin{equation}
ds^2_{10}\,=\,g_s \alpha' Ne^{ \phi}\,\Big[\,\frac{dx^2_{1,2}}{g_s \alpha' N}\,+\,e^{2h}\,\big(\,d\theta^2+\sin^2\theta d\phi^2\,\big)\,+\,
d\rho^2\,+\,{1\over 4}\,\sum_{i}(w^i-A^i)^2\,\Big] +e^{- \phi} dz^2 \,\,, 
\label{metric1}
\end{equation}
where $\phi$ is the type IIB dilaton and $h$ is a function which depends on the dimensionless radial coordinate $\rho$:
\begin{equation}
e^{2h}\,=\,\rho \coth 2\rho\,-\,{\rho^2\over \sinh^2 2\rho}\,-\,
{1\over 4}\,\,,\;\;\;\;\;\;\qquad \qquad
e^{-2\phi}\,=\,e^{-2\phi_0}{2e^h\over \sinh 2\rho}\,\,.
\label{MNsol}
\end{equation}
The one-forms $A^i$ encode a fibration of the two-sphere $(\theta,\phi)$ over the three-sphere parameterized by the set of $su(2)$ left-invariant one-forms $w^i$, $(i=1,2,3)$. The type IIA background also includes a RR potential $C^{(3)}=C^{(2)} \wedge dz$, where $C^{(2)}$ is the original RR potential of the MN solution. 

One unsatisfactory aspect of the picture of the Yang-Mills theory on the cylinder in the type IIA MN set-up could be the bad behavior of the dilaton at large values of the radial coordinate. This signals the decompactification of the eleventh dimension of M-theory. Under this perspective, it is quite natural to up-lift the MN solution to eleven dimensional supergravity.  One gets:
\begin{eqnarray}  \label{11metric}
\nonumber ds^2_{11}&=& e^{2/3\phi}\left[ dx_{1,2}^2+\alpha'g_sN e^{2h}(d\theta^2+\sin^2\theta d\phi^2)+\alpha'g_sNd\rho^2+\frac{\alpha'g_sN}{4}\sum_i(\omega^i-A^i)^2\right]\,+\,\\ &&\ \ +\, e^{2/3\phi}dy^2~+\ e^{-4/3\phi}dz^2,\\ C^{(3)}&=&C^{(2)}\wedge dz,\nonumber\end{eqnarray} 
where $C^{(3)}$ is the magnetic potential under which the M5-branes are charged and the new eleventh coordinate $y$ is also periodic. 
 
This solution corresponds to have $N$ M5 branes wrapping, besides an $S^2$, the $y$ circle\footnote{This $S^2$ is a mixture of the two two-spheres of the geometry, see \cite{Bertolini:2002yr}.} and smeared in the $z$ direction. The gauge theory we are describing corresponds thus to the worldvolume theory of such M5-branes. We begin noticing an intriguing property of such description. Starting from eleven dimensions, for small $\rho$ it is fine to go down to IIA along the $y$-circle: it has small radius and the IIA theory is well behaved. Thus we get a solution in terms of D4-branes. For big $\rho$ instead the radius of the $y$-circle becomes big. If we insist on making the dimensional reduction along that circle, the dilaton diverges. But, as it is clear from eq. (\ref{11metric}), the radius of the $z$-circle becomes small in the large $\rho$ limit: it is now possible to reduce along $z$ and get a well behaved IIA solution in terms of NS5-branes. 
The same happens in the type IIB MN solution, where for small $\rho$ one has a well behaved solution in terms of D5 branes, while for big $\rho$ the dilaton diverges and one needs to S-dualize the background and gets instead NS5 branes. 

We see here that the eleven dimensional picture gives a unifying picture of this phenomenon in terms of a unique theory on the M5-brane worldvolume. 

\section{M5-brane worldvolume theory}  \label{M5}
Let us consider a M5-brane where its worldvolume geometry is of the form ${\mathbb{R}}^{1,2}\times S^1\times \Omega^2$, where $\Omega^2$ is a two dimensional compact manifold whose volume element is $V_2d\Omega_2$ ($\int d\Omega_2=4\pi$). $S^1$ is a circle of radius $g_s\sqrt{\alpha'}$. Standard Kaluza-Klein reduction on the internal three dimensional manifold ${\bf{S}}\,=\,S^1\times\Omega^2$ gives rise to the three dimensional gauge theory. To study such theory we need to investigate the M5 worldvolume dynamics. Our main tool will be the (covariant) worldvolume action written in \cite{Pasti:1997gx}. In such formalism (usually called PST formalism), it is better if the internal manifold contains a factorized circle (or more generally its first Betti number should be different from zero)\footnote{This is related to the fact that the PST scalar is naturally an angular variable.}.

The M5 worldvolume PST action is \cite{Pasti:1997gx}:
\begin{equation}
\label{PST}
S=T_{M5}\int d^6\xi\left(-\sqrt{-det(g+\tilde{H})}+\frac{\sqrt{-det g}}{4\partial a\cdot\partial a}\partial_ia(\star H)^{ijk}H_{jkl}\partial^la\right)
+\frac{T_{M5}}{2}\int F\wedge C^{(3)}\, ,
\end{equation}
where $g$ is the pullback of the eleven dimensional background metric, $a$ is the PST scalar and the three-form $H$ is defined as
\begin{equation}
 H=F-C^{(3)},
 \end{equation} 
 where $F$ is a worldvolume three-form field strength ($F=dA_{(2)}$).
The two-form $\tilde{H}$ is defined as 
\begin{equation}
\label{Htilde}\tilde{H}^{ij}~=~\frac{1}{3!\sqrt{-detg}}\frac{1}{\sqrt{-(\partial a)^2}}\epsilon^{ijklmn}\partial_kaH_{lmn},
\end{equation}
and $T_{M5}$ is the M5-brane tension. 

Specifying to our case, we want to implement the KK reduction on the internal manifold ${\bf S}$. We need to do a gauge fixing for the PST scalar. The most natural one is $a~=~y$, where $y\in [0,2\pi \sqrt{\alpha'}g_s]$ parameterizes the $S^1$. Consistently with this choice, we consider the following two-form worldvolume potential $A_{(2)}$:
 \begin{equation} \label{AA}
 \frac{A_{(2)}}{(2\pi)^2g_s}=\alpha'\frac{y}{2\pi g_s}\frac{1}{2} F_{ab}dx^a\wedge dx^b+\alpha'^{\frac{3}{2}}\Sigma d\Omega_2 \,\, ,
 \end{equation}   
where the indices $a,\, b$ span the three dimensional Minkowski space ($a,\ b=0,1,2$), $F_{ab}=\partial_aC_b-\partial_bC_a$, $C_a$ is a three dimensional vector and  $\Sigma$ is a dimensionless scalar field depending on the coordinates $x_a$. 
Even if the properties of the quantum M5-brane theory are subtle, it is quite natural to quantize the possible variations of the two-form integrated on the (contractible) two-cycle. One way to see this is to use dualities that relate the two form $A_{(2)}$ to the NS-NS $B_{(2)}$ field. The relevant quantity which is allowed to change by integer units is:
\begin{equation}
\frac{1}{(2\pi)^3g_s\alpha'^{3/2}}\int A_{(2)}.
\end{equation}
The field $\Sigma$ is thus naturally periodic. With our normalizations (\ref{AA}) its period is:
\begin{equation} \label{TSigma} 
T_{\Sigma}\,=\,\frac{1}{2}\,\,.
\end{equation}
Inserting the ansatz (\ref{AA}) in (\ref{PST}) one gets the three dimensional action. Expanding it in powers of $\alpha'$ and discarding the constant term, the leading term that we get is the three dimensional flat spacetime action:
\begin{equation} 
S=-\left[\frac{1}{2}\frac{2\pi\sqrt{\alpha'} g_s}{V_2}\int d^{1,2}x\partial_c\Sigma\partial^c\Sigma+\frac{1}{2}\int d^{1,2}x\epsilon^{abc}F_{ab}\partial_c\Sigma\right]\, . \label{PST3}
\end{equation}
Thanks to the second term in (\ref{PST3})  $\Sigma$ is naturally interpreted as the scalar field dual to the vector one. 
We can interpret the vector field $C_a$ as a Lagrange multiplier and to vary with respect to it, enforcing the Bianchi identity constraint on $\Sigma$. Otherwise we can vary with respect to the vector $\partial_a\Sigma$ and get the standard action in terms of the vector field $C_a$. 
As a result, it is quite obvious to relate the volume $V_2$ of the two cycle to the square of the inverse of the three dimensional gauge coupling constant in the following way:
\begin{equation}
g_{YM3}^2\,=\,\frac{2\pi\sqrt{\alpha'} g_s}{V_2}. 
\end{equation}
It is easy now to recognize (\ref{PST3}) as the standard free Maxwell action in three dimensions.

Notice that the dualization term, which allows us to identify the scalar $\Sigma$ as the dual to the vector field, it is contained here automatically in the worldvolume theory. Such theory is moreover written in terms of the scalar $\Sigma$, appearing in (\ref{PST3}) the kinetic term for this field .

\section{${\cal N}=1$ super Yang-Mills on the cylinder} 
\label{supot}

We are now ready to interpret the eleven dimensional solution given in section \ref{sec11} from the field theory point of view. This is a ${\cal N}=2$ supersymmetric field theory in three dimensions obtained from a circle reduction of ${\cal N}=1$ (pure) super Yang-Mills in four dimensions. Being the M5-branes smeared in the $z$-circle, the gauge group is $U(1)^{N-1}$ (the degree of freedom corresponding to the center of mass of the system is decoupled). Applying the analysis we described in the previous section (but omitting the dualization term), namely making an M5-probe computation and expanding the result in powers of $\alpha'$, for the {\it i}-th gauge group we get:  
\begin{equation}
S_i=-\frac{1}{2}\int d^3\xi\left[ \frac{2\pi}{g^2_{YM4}R}\partial_ab_i\,\partial^ab_i\,+\, \frac{g^2_{YM4}}{2\pi R}\partial_a\Sigma_i\partial^a\Sigma_i\right]\,\, ,
\end{equation}
where $\frac{1}{g^2_{YM4}}~=~\frac{N}{4\pi^2}\rho \tanh{\rho}$ and we are restricting to the case $\theta_{YM}=0$. The periodic scalars $b_i$ correspond to fluctuations in the $z$ direction and are given by $z=2\pi\sqrt{\alpha'}\,b$. Redefining the field $\Sigma_i$ as 
\begin{equation}\label{gamma} 
\gamma_i\,=\,\frac{g^2_{YM}}{2\pi}\Sigma_i,
\end{equation}
it is easy to write the action in terms of the holomorphic field $\Psi_i\,=\,b_i\,+\,i\gamma_i$ simply as:
\begin{equation} 
S_i\,=\,-\frac{\pi}{g^2_{YM}R}\int d^3x\,\partial_a\Psi_i\partial^a\Psi_i\,\, .
\end{equation}
From (\ref{TSigma}) and (\ref{gamma}) we can read off the period of $\gamma_i$:
\begin{equation}\label{Tgamma}
T_{\gamma}\,=\,\frac{g^2_{YM}}{4\pi}.
\end{equation}

We look now for some M-brane configuration generating a superpotential. For this to happen one has to check that in presence of such M-branes there are two fermionic zero-modes \cite{Witten:1996bn}.
The superpotential they generate is \cite{Witten:1996bn}:
\begin{equation} \label{supotM2}
W\,\sim\,\mu^3\sum_i{\rm e}^{i\,S_{M_i}},
\end{equation}
where $\mu$ is a dimensionful scale (with inverse length dimension) related to the value of the radial variable at which the computation is made (it is the same value as the one at which the actions $S_{M_i}$ are evaluated). 

To have some intuition on what are these configurations, we start noticing that an instanton configuration is an (Euclidean) M2-brane wrapped along the $\Omega_2$ introduced in section \ref{M5} and the entire $z$ circle. In presence of such configuration, from the index theorem, we expect to have $2N$ fermionic zero-modes. 
Before doing the zero-mode counting, we have to remember that along the $z$-circle there are $N$ M5-branes. It is well known that an M2-brane can end on an M5-one. Analogously, the instantonic M2-brane can open itself and end on one of the $N$ M5s. We are thus led to consider $N$ objects that are more basic than the instantonic M2-brane and can be seen as its constituents. These objects are the $N$ M2-branes stretching between two consecutive M5-branes. 
In this way we get a picture very close to the field theoretical one of the instanton as being composed by more fundamental instanton partons \cite{Belavin:1979fb}, the so-called "fractional instantons'' (our open M2-branes). 

For the kappa-symmetry analysis of this kind of configurations one can show that half of the supersymmetries of the background are preserved (in the proper limit) by such M2-branes \cite{Canoura:2007gw}. This implies that two (real) supersymmetries are broken and, consequently, there are two fermionic zero modes in this background: they are the goldstinos of the broken supersymmetries. The equations of motion for the fermionic fluctuations (up to second order in fermions) in an arbitrary bosonic background were written in \cite{Mart}. A direct inspection of such equations shows that the presence of other zero modes is unlikely. Therefore, we assume that in this background there are just the two goldstinos zero modes we discussed here.

\subsection{The non-perturbative superpotential}

To proceed further and write explicitly the superpotential generated by these M2-branes via the formula (\ref{supotM2}), we need the form of their worldvolume action. As we are considering open M2-branes stretching between two M5s, we have to pay attention to the fact that on the M5-brane worldvolume, the boundary of an M2-brane (a string) sources a potential. This is 
precisely the $A_{(2)}$ M5 worldvolume two-form potential. The coupling of open M2 to it is easily evaluated \cite{Strominger:1995ac} (perhaps the best way of seeing it is by requiring gauge invariance for the $C^{(3)}$ potential). The resulting (open) M2 worldvolume action is:
\begin{equation}\label{M2}
iS_{M2_i}\,=\,-T_{M2}\int d^3\xi\sqrt{det\, g}\,+\, i\,T_{M2}\int\left( C^{(3)}-F\right),
\end{equation}
where $T_{M2}$ is the tension of the M2-brane.

Out of the $N$ M5-branes we need to decouple the center of mass. To this aim, we consider one (non dynamical) M5-brane fixed at $z=0$. For the {\it i}-th M2-brane (extending between the $({\it{i}}-1)$ and the {\it i} M5-branes), we can evaluate the action (\ref{M2}):
 \begin{equation}
 iS_{M2_i}\,=\,-\frac{8\pi^2}{g^2_{YM}}\left[(b_i-b_{i-1})\,+\,i\,\frac{g^2_{YM}}{2\pi}(\Sigma_i-\Sigma_{i-1})\right]\,=\,-\frac{8\pi^2}{g^2_{YM}}\left(\Psi_i-\Psi_{i-1}\right)\,=\,-\frac{8\pi^2}{g^2_{YM}}(\Delta\Psi)_i \,\, , 
\end{equation}
 where we make again the computation at $\theta_{YM}=0$. We must pay special attention to the {\it N}-th M2-brane, the one extending between the $N-1$ M5-brane and the non dynamical one. In this case its action is not independent of the others, but it is given by the difference between the instantonic one (the one corresponding to the M5 extending along the entire circle $z$) and all the others $N-1$. We call it Kaluza-Klein monopole (as it has been named the analogous configuration in field theory in \cite{Davies:1999uw}). We easily compute its action and find perfect agreement with field theory (see eq. (2.10) of \cite{Davies:1999uw}):
 \begin{equation}
 S_{KK}\,=\,-\frac{8\pi^2}{g^2_{YM}}\frac{R}{\sqrt{\alpha'}}\,+\, \sum_{i=1}^{N-1}\frac{8\pi^2}{g^2_{YM}}(\Delta\Psi)_i.
 \end{equation}
 
Putting all together and redefining $(\Delta\Psi)_i$ as $\Phi_i$, we get the superpotential (\ref{supotM2}):
 \begin{equation}\label{M2supot}
 W\,=\,M^3\left(\sum_{i=1}^{N-1}{\rm e}^{-\frac{8\pi^2}{g^2_{YM}}\Phi_i}+{\rm e}^{-\frac{8\pi^2}{g^2_{YM}}\frac{R}{\sqrt\alpha'}+\sum_{i=1}^{N-1}\frac{8\pi^2}{g^2_{YM}}\Phi_i}\right).
 \end{equation}
This nicely reproduces the field theory one \cite{Davies:1999uw}. By extremizing it we get the M-branes equilibrium configuration:
\begin{eqnarray}    \label{vacua}
&&\langle\Phi_i\rangle\,=\,\frac{R}{N\sqrt{\alpha'}}\,+\,i\,\frac{g^2_{YM}}{4\pi}\,\frac{ k}{N} \qquad \qquad k \,\, \in \,\, {\mathbb {Z}} \,\, ,\nonumber \\
&& \langle W \rangle\,=\, N\Lambda^3,
\end{eqnarray}
where $k$ is defined modulo N (see (\ref{Tgamma})) and labels the $N$ vacua resulting from the breaking of the ${\mathbb {Z}}_{2N}$ R-symmetry to ${\mathbb {Z}}_2$. These vacua are related to the ones of gaugino condensation. Domain walls naturally follow.
Via this superpotential, the magnetic photons do get a mass. This is a signal of  confinement. 

\section*{Acknowledgments}
I am pleased to thank Paolo Merlatti for delightful collaborations leading to the results presented in this talk. This work was partially supported by MCyT and FEDER under grant FPA2005-00188, by Xunta de Galicia (Conseller\'\i a de Educaci\'on and grant PGIDIT06PXIB206185PR) and by the EC Commission under grant  MRTN-CT-2004-005104.

\end{document}